\documentclass[showpacs,aps,prx,twocolumn,groupedaddress,longbibliography]{revtex4-1}

\usepackage{threeparttable}
\usepackage{graphicx}
\usepackage{amsmath}

\begin{document}


\title{Quantum information processing using quasiclassical electromagnetic interactions between qubits and electrical resonators}


\author{Andrew J. Kerman}
\affiliation{MIT Lincoln Laboratory, Lexington, MA 02420}


\date{\today}

\begin{abstract}
Electrical resonators are widely used in quantum information processing, by engineering an electromagnetic interaction with qubits based on real or virtual exchange of microwave photons. This interaction relies on strong coupling between the qubits' transition dipole moments and the vacuum fluctuations of the resonator in the same manner as cavity QED, and has consequently come to be called ``circuit QED" (cQED). Great strides in the control of quantum information have already been made experimentally using this idea. However, the central role played by photon exchange induced by quantum fluctuations in cQED does result in some characteristic limitations. In this paper, we discuss an alternative method for coupling qubits electromagnetically via a resonator, in which no photons are exchanged, and where the resonator need not have strong quantum fluctuations. Instead, the interaction can be viewed in terms of classical, effective ``forces" exerted by the qubits on the resonator, and the resulting resonator dynamics used to produce qubit entanglement are purely classical in nature. We show how this type of interaction is similar to that encountered in the manipulation of atomic ion qubits, and we exploit this analogy to construct two-qubit entangling operations that are largely insensitive to thermal or other noise in the resonator, and to its quality factor. These operations are also extensible to larger numbers of qubits, allowing interactions to be selectively generated among any desired subset of those coupled to a single resonator. Our proposal is potentially applicable to a variety of physical qubit modalities, including superconducting and semiconducting solid-state qubits, trapped molecular ions, and possibly even electron spins in solids.
\end{abstract}


\maketitle

\section{Introduction}\label{s:intro}

Microwave electrical resonators are already an important tool for manipulating quantum information using electromagnetic interactions \cite{nori_review}. As a means of quantum information storage or communication, they are used or proposed in a variety of schemes involving, for example, trapped molecular ions \cite{molions}, neutral polar molecules \cite{polar}, Rydberg atoms \cite{wallraffryd,*molmerryd,*schmiedryd,*lukinryd,haroche}, superconducting Josephson-junction qubits \cite{blaisarch,*schusterCQED,wallraffCQED,*noriCQED}, and electron spins in solids \cite{schuster,rareearth}. In nearly all of these cases, the resonator is used in a way familiar from optical cavity quantum electrodynamics \cite{haroche,ocQED}, in which the qubits exchange real or virtual photons with the cavity, and where the figure of merit is the speed with which this exchange occurs (or equivalently, the strength of the coherent coupling between the qubit's transition dipole moment and the resonator's vacuum fluctuations). In fact, many of the seminal cavity QED results have been replicated in a circuit environment, a field now known as circuit QED \cite{blaisarch,*schusterCQED,wallraffCQED,*noriCQED} (cQED).

Just as with its optical predecessor, cQED for a single qubit interacting with a resonator can be approximately described by the Jaynes-Cummings model \cite{*[{See, for example: }] [{.}] Prep}, with the Hamiltonian \cite{blaisarch,*schusterCQED}:

\begin{equation}
\hat{H}_{JC}=\frac{\hbar\omega_q}{2}\hat\sigma_q^z+\hbar\omega_r\left(\hat{a}^\dagger \hat{a}+\frac{1}{2}\right)+\hat{H}_\perp
\end{equation}

\noindent where $\hbar\omega_q$ is the energy splitting between the qubit's two states $|e\rangle$ and $|g\rangle$, whose Hilbert space is acted on by $\hat\sigma_q^z$, $\hat{a}$ is the annihilation operator for photons of a resonator mode with frequency $\omega_r$, and the transverse interaction is given in the rotating-wave approximation by:

\begin{equation}
\hat{H}_\perp\approx\hbar g_\perp\left( \hat{a}\hat\sigma_q^+ + \hat{a}^\dagger\hat\sigma_q^- \right)\label{eq:Hperp}
\end{equation}

\noindent where $\hat\sigma_q^\pm\equiv(\hat\sigma_q^x\pm i\hat\sigma_q^y)/2$ are the qubit raising and lowering operators. The coupling strength has the generic form: $\hbar g_\perp=\mu_q\delta F_{rms}$, where $\mu_q\equiv\langle g|\hat\mu_q|e\rangle$ is the qubit's transition dipole moment ($\hat\mu_q$ is the appropriate electric or magnetic dipole moment operator for the qubit), and $\delta F_{rms}$ is the rms amplitude of the resonator's electric or magnetic field vacuum fluctuations. Equation~\ref{eq:Hperp} then describes an exchange of photons between the qubit and the resonator mode, via interaction between the resonator's vacuum field and the qubit's dipole moment.

If the detuning between resonator and qubit $\Delta_q\equiv\omega_r-\omega_q$ is much smaller than $g_\perp$, Rabi flopping at frequency $g_\perp$ occurs in which a single photon is exchanged between them (known in the frequency domain as the vacuum Rabi splitting), and with it one bit of quantum information, which can then be transferred to another qubit coupled to the same resonator \cite{silanpaa}. Achieving high fidelity in this transfer, however, requires extremely high $Q$ for the resonator, since the photon spends a time of order the Rabi period stored inside it. To relax this requirement, most cQED is performed instead in the regime of large detuning $\Delta_q\gg g_\perp$, known as the dispersive limit, where the effective coupling is second order in $\hat{H}_\perp$, with a magnitude $\approx g_\perp^2/\Delta_q$. Although $g_\perp$ must be made larger by a factor $\Delta_q/g_\perp\gg1$ compared to the resonant case to achieve the same operation speed, the time spent by the photon in the resonator is effectively reduced to $\Delta_q^{-1}$; this can be viewed as virtual photon exchange, in the sense that the photon moves to a different energy by entering the resonator, but only for a time consistent with the uncertainty principle. The detuning also provides a natural means of controlling the effective qubit-resonator interaction, since in most cases the transition dipole $\mu_q$ of the qubit is not dynamically adjustable \cite{*[{An exception is in: }] [{ where the qubit's intrinsic electromagnetic coupling can be adjusted independently of the cavity.}] tunableQED}, whereas its energy splitting $\hbar\omega_{qb}$ often is \cite{*[{An alternative method for achieveing tunability is to keep the detuning fixed at a large value and use the resulting weak, static coupling in conjunction with a strong driving field, e.g.: }] [{ }] IBM_crossres,*wallraff_sideband,*IBM_sideband}.

This cQED paradigm has been extremely fruitful for quantum information processing, and has already been used to demonstrate many important QIP functions including complete multi-qubit algorithms \cite{martinisALG,*reedALG}. In some cases it is already the basis for envisioned scaling to much larger systems \cite{helmer,*rezQu,*divincenzo}. In spite of its great success in the QIP area, however, cQED also has some characteristic limitations when used in this context. For example, when a qubit is engineered to have a large transition dipole so that it can have strong $g_\perp$ for cQED, it necessarily becomes more sensitive to its electromagnetic environment (i.e. it can couple to spurious environmental modes as well as the desired resonator mode) \cite{MRFS}. In addition, since cQED effectively uses quantum states of the resonator to store or transport quantum information, its protocols are quite sensitive to the presence of spurious photon populations in the resonator (including in some cases those in higher modes) which are often encountered in these experiments \cite{sears,multimode,rigetti}, and also to a lesser extent its quality factor $Q$. Next, although adjustment of the detuning $\Delta_q$ does provide effective control of the relevant interactions in experiments to date, it does not in general allow a very strong suppression of the coupling when it is intended to be off, since it scales only as $\sim1/\Delta_q$; furthermore, the adjustability of the qubit energy required for this control necessarily implies that the qubits are sensitive to noise in whatever parameter is used for this adjustment (e.g. charge or flux noise, or noise in the external bias itself) \cite{IBM_crossres,*wallraff_sideband,*IBM_sideband}. Finally, since in the dispersive limit of cQED interaction with a qubit requires exchanging a virtual photon with it, direct $N$-qubit interactions get exponentially weaker as $N$ increases. This implies that sequential, pairwise interaction between qubits will likely remain the best way to achieve $N$-qubit entanglement in cQED, even when all of the $N$ qubits are coupled to the same resonator. In this case one must take care when scheduling the various ramps of the qubit energies, since whenever any two qubits come close to resonance with each other they must both be very far detuned from the resonator to avoid spurious entanglement.

In this paper, we describe an alternative approach for coupling qubits and electrical resonators via electromagnetic interactions, which does not involve any photon exchange between them, and therefore largely avoids these limitations. In contrast to the transverse interaction $\hat{H}_\perp$ of cQED [c.f., eq.~\ref{eq:Hperp}], we will show how one can realize a ``longitudinal" interaction of the form:

\begin{equation}
\hat{H}_{||}\approx\hbar g_{||}(\hat{u}\hat\sigma_q^z)\label{eq:Hpar}
\end{equation}

\noindent where $\hat{u}\equiv(\hat{a}+\hat{a}^\dagger)/\sqrt{2}$ is the dimensionless resonator coordinate. This interaction energy, being linearly proportional to the ``position" of the resonator, constitutes an effective force acting on it which depends on the qubit's internal state (via the operator $\hat\sigma_q^z$). Our proposal is built on an analogy between this interaction and demonstrated methods \cite{Znature,haljan} developed by M\o lmer and S\o rensen \cite{MSPRL,*MSPRA} and Milburn \cite{milburn} for entangling trapped atomic ions via qubit-state dependent forces acting on their collective center-of-mass vibrational modes \cite{*[{Related methods specific to Josephson-junction-based ``flux" qubits were described in: }] [{ and: }] nori,*semba} (whose role is played here by the microwave resonator mode). In contrast to cQED as described above, our scheme involves only quasi-classical resonator states (whose vacuum fluctuations can be small), under the influence of effectively classical, qubit-state-dependent forces \cite{dispersive}. As a result, the resonator dynamics which drive the gate operations are classical and macroscopic in nature, do not depend at all on the qubit frequency $\omega_q$, and are insensitive to thermal or other fluctuations and damping. In addition, because the interaction does not rely on photon exchange, all of the qubits coupled to the same resonator can be entangled in the same amount of time it takes to entangle only two of them, and any subset of the qubits can be similarly entangled with negligible effects on the others.

In section~\ref{s:system} below, we describe the general qubit-resonator system under consideration. Section~\ref{s:gates} contains a detailed description of how entangling operations can be achieved between qubits coupled to a common resonator, without any real or virtual exchange of photons. In section~\ref{s:errors} we consider in detail the leading sources of error in two-qubit entangling operations, and evaluate these errors for a variety of different physical qubit modalities, which are tabulated in table~\ref{tab:sample}. We conclude in section~\ref{s:conclusion} with a summary of the differences between our proposal and cQED. Appendix~\ref{a:quasicharge} contains conceptual details about the classical interpretation of the forces used to implement our gates, appendices~\ref{a:errors} and~\ref{a:hm} describe certain aspects of the gate error calculations, appendix~\ref{a:modalities} the assumptions and parameter values used to evaluate the gate errors shown in table~\ref{tab:sample}, and appendix~\ref{a:ioncomp} a comparison between our proposal and the analogous trapped-ion gates.

\begin{figure*}
\includegraphics[width=6.0in]{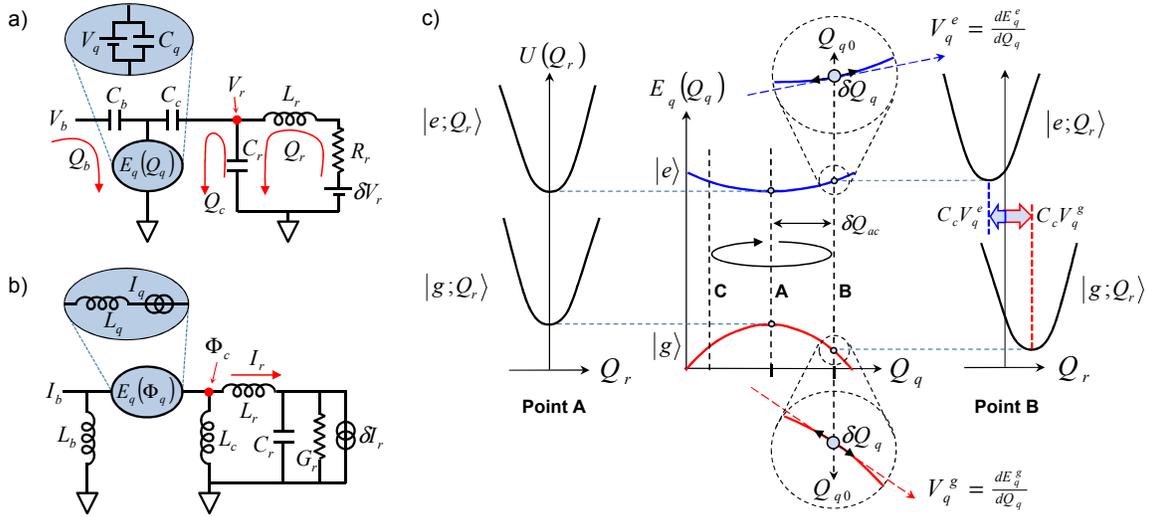}
\caption{\label{fig:system} Electromagnetic qubit/resonator coupling without exchange of photons. (a) shows the electric circuit analyzed in detail; (b) is its exact dual \cite{likharevdual,*kadin,*mooijfluxchg,AJKdual,likharevbloch,*hekking}, governed by equations of identical form, via the transformation: $Q\leftrightarrow \Phi,C\leftrightarrow L,V\leftrightarrow I,Y\leftrightarrow Z$ \cite{dividers}, in which loop charges [red arrows in (a)] become node fluxes [red dots in (b)] \cite{devoret}. Resonator damping, and the associated fluctuations, are modeled in (a) via $R_r$ and the Langevin noise source $\delta V_r$, and in (b) by $G_r$ and $\delta I_r$. The qubit in each case is described by a state-dependent classical potential energy $E_q(p)$, with $p\in Q_q,\Phi_q$, an example of which is shown in (c) for the electric case $p=Q_q$ by red and blue lines for $|g\rangle$ and $|e\rangle$, respectively. In eq.~\ref{eq:expand}, we expand $E_q(Q_q)$ in excursions $\delta Q_q$ about a quasi-static bias point $Q_{q0}$, as illustrated in (c). This expansion can be described by the circuit elements in the shaded ovals of (a) and (b). The leading (linear) term results in an effective force on the resonator which displaces its equilibrium position by a (qubit-state-dependent) amount $C_cV_q$ [c.f., eq.~\ref{eq:stillU}], as illustrated in (c) for two different qubit bias points A and B: point A is a so-called ``degeneracy" point where $V_q^\pm=0$, and there is no force or displacement, while B illustrates the nonzero case. In our proposed multiqubit gates, we modulate the force by modulating the qubit bias point, an example of which is illustrated in (c) by the curved arrow going from A$\rightarrow$B$\rightarrow$C$\rightarrow$A, and so on. }
\end{figure*}

\section{General qubit-resonator system}\label{s:system}

\begin{figure}
\includegraphics[width=3.4in]{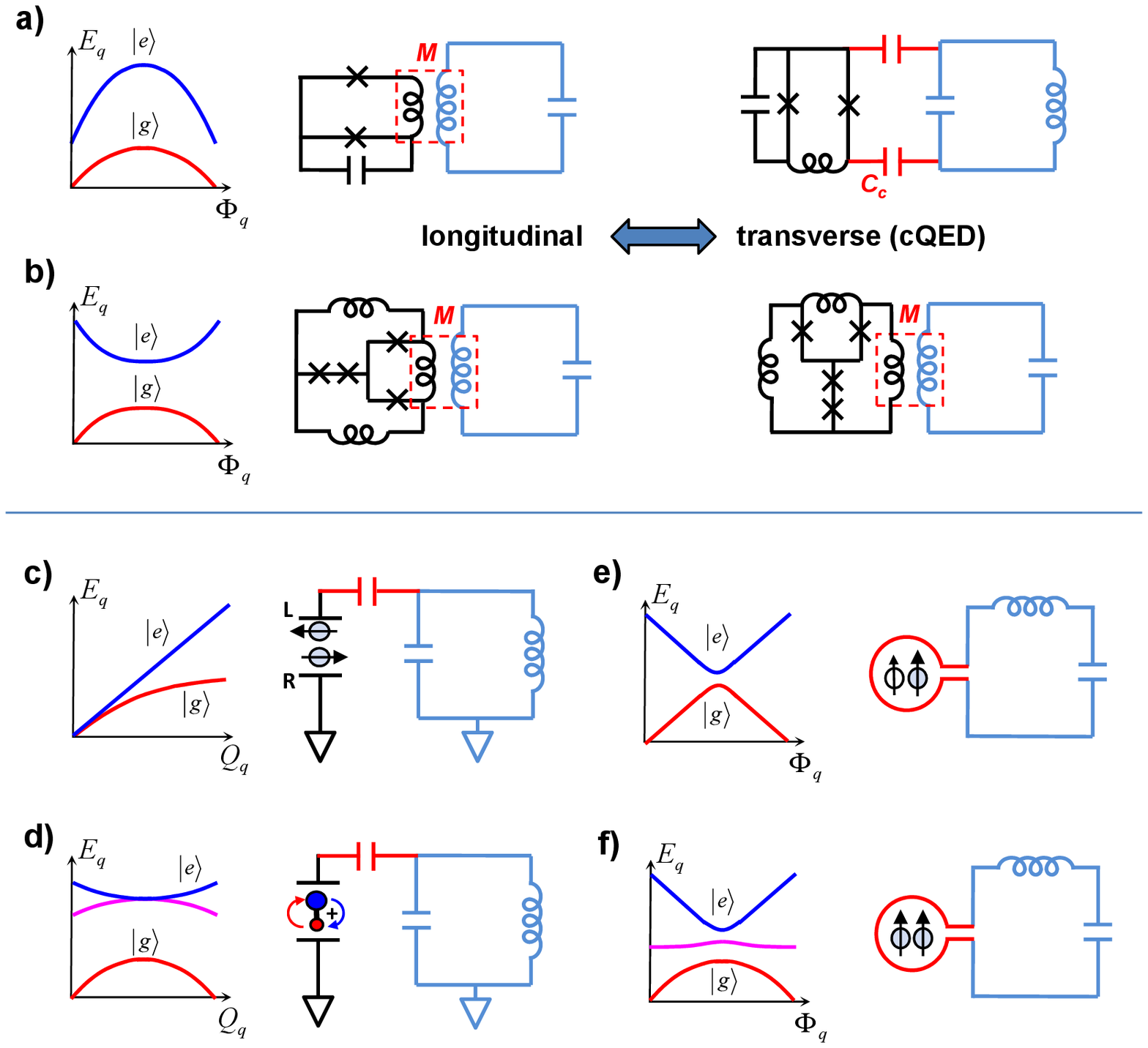}
\caption{\label{fig:examples} Examples of physical qubits coupled to resonators in the manner of our proposal. For each case we show a schematic of the equivalent circuit with the qubit in black, resonator in light blue, and the coupling element in red (note that the qubit bias connections are not shown). Next to each case are the corresponding energies $E_g(p)$ and $E_e(p)$ [see appendix~\ref{a:modalities} for details]. (a) and (b) are the transmon \cite{blaisarch,*schusterCQED,wallraffCQED,*noriCQED} and flux qubit \cite{mooij,*MITtunable,orlando}, respectively, both of which have been extensively demonstrated in a cQED architecture; the corresponding cQED circuits are shown for comparison on the right side of these panels. The inductors inside the qubits represent the usual geometrical loop inductances, which are typically neglected since they are much smaller than the Josephson inductances of the JJs. They are shown here only to illustrate mutual inductive coupling between qubit and resonator. (c)-(f) show how our scheme can also be applied to a number of other qubit modalities: (c) singlet-triplet double quantum dot \cite{petta,*laird,spinQDs2,QDcouple}; (d) polar molecular ion \cite{molions}, where the third curve (shown in magenta) corresponds to the two degenerate $J=1,m_J=\pm1$ levels; (e) $^{31}$P coupled electron and nuclear spins in $^{28}$Si \cite{Si}, where we show only one nuclear spin orientation. The splitting at zero field is due to hyperfine coupling; (f) Nitrogen-vacancy center in diamond \cite{balaNV}, with a fixed transverse magnetic field, as a function of an additional longitudinal field. The third (magenta) curve shows the two degenerate levels which become $m_S=0$ in the limit of large longitudinal field. }
\end{figure}

We first consider a single qubit coupled to a microwave resonator, using the circuits of figs.~\ref{fig:system}(a) and (b) for capacitive and inductive qubit/resonator coupling. These two circuits are chosen to be exactly \textit{dual} \cite{likharevdual,*kadin,*mooijfluxchg,AJKdual,likharevbloch,*hekking} to each other, so that the solution for one case can be mapped directly to the other using the transformation: $Q\leftrightarrow \Phi,C\leftrightarrow L,V\leftrightarrow I,Y\leftrightarrow Z$. In each of these circuits, the qubit is described by a purely classical (though state-dependent) potential energy $E_q(p)$, where $p$ is a classical parameter in the qubit Hamiltonian on which the energies of its eigenstates $|g\rangle$ and $|e\rangle$ depend. In the figure $p$ is either the induced charge on a qubit gate electrode $Q_q$ [fig.~\ref{fig:system}(a)], or the magnetic flux through a qubit loop $\Phi_q$ [fig.~\ref{fig:system}(b)]; the former case is illustrated schematically in fig.~\ref{fig:system}(c) by the solid red and blue lines for the qubit states $|g\rangle$ and $|e\rangle$, respectively. In the absence of coupling between qubit and resonator, $p$ is determined only by the external bias of the qubit ($V_b$ or $I_b)$. With the coupling turned on, which for the circuits in figs.~\ref{fig:system}(a) and (b) corresponds to nonzero $C_c$ and $L_c$, respectively, $p$ acquires a contribution induced by the resonator. In treating the qubit as a purely classical potential energy $E_q(p)$, we are excluding by construction any cQED-like transverse interaction [c.f., eq.~\ref{eq:Hperp}] by which the qubit can exchange a photon with the resonator and change its internal state. Although for implementation of our scheme it is preferable that there is simply no such interaction present at all in the system, it can still work well in systems where a nonzero transverse interaction is present, but is made negligible by using large $\Delta_q$. Several examples are shown schematically in fig.~\ref{fig:examples}, for a variety of different physical qubit modalities, spanning both of these cases (see appendix~\ref{a:modalities} for details).

We now analyze our system in detail, considering specifically the capacitively-coupled circuit of fig. \ref{fig:system}(a) for definiteness. We write its classical potential energy in terms of the loop charges $Q_b,Q_c,Q_r$ \cite{devoret} shown in the figure (for the moment taking the damping resistance $R_r$ and associated noise fluctuations $\delta V_r$ to be zero):

\begin{eqnarray}
U(Q_r,Q_c,Q_b)=E_q(0)&+&\frac{(Q_r-Q_c)^2}{2C_r}+\frac{Q_c^2}{2C_c}+\frac{Q_b^2}{2C_b}\nonumber\\
+\underbrace{\int_0^{Q_q} V_q(Q_q^\prime) dQ_q^\prime}_{E_q(Q_q)-E_q(0)}&-&\int_0^{Q_b} V_b(Q_b^\prime)dQ_b^\prime\label{eq:Etot}
\end{eqnarray}

\noindent where $Q_q\equiv Q_b+Q_c$, $V_q(Q_q)\equiv dE_q/dQ_q$ is the voltage across the qubit, $V_b(Q_b)$ is defined as the bias voltage for which the source has supplied a charge $Q_b$. The terms in the second line are the potential energies of the qubit and the source, respectively, and can be viewed as work done by (or on) the source as $V_b$ is turned up from zero.

We now seek to expand eq.~\ref{eq:Etot} in deviations of the charges $Q_n$ ($n\in b,c,r$) about their minimum-energy, quasistatic solutions $Q_{n0}$, which are given formally by: $Q_{b0}=C_b(V_b-V_q),\;Q_{c0}=-C_c V_q,\;Q_{r0}=Q_{c0}$ (these solutions must be obtained self-consistently since $V_q$ depends implicitly on $Q_q$). Note that total energy conservation implies that $U$ evaluated at this point: $U(Q_{r0},Q_{c0},Q_{b0})$ will be independent of $V_b$, a fact which can be verified using eq.~\ref{eq:Etot}. We now expand the qubit energy about its static bias point, writing: $Q_n\equiv Q_{n0}+\delta Q_n$:

\begin{equation}
E_q(Q_{q0}+\delta Q_q)\approx E_q(Q_{q0})+V_q\delta Q_q+\frac{\delta Q_q^2}{2C_q}+...\label{eq:expand}
\end{equation}

\noindent where the linear term is proportional to the qubit voltage $V_q$, and the quadratic term can be written in terms of an effective dynamic capacitance of the qubit $C_q\equiv (d^2E_q/dQ_q^2)^{-1}$ (also known as ``quantum capacitance" \cite{LQmooij,*CQdelsing,*CQhakonen}), both of which are evaluated at $Q_{q0}$. As illustrated in fig.~\ref{fig:system}(a), this expansion allows us to view the potential energy of the qubit in a circuit context as a voltage source $V_q$ in parallel with the capacitance $C_q$ (in the dual magnetic case of fig.~\ref{fig:system}(b) this becomes a series current source $I_q$ and inductance $L_q$). We emphasize that $V_q$ and $C_q$ depend on both the internal state of the qubit, and its bias $Q_{q0}$, though we have suppressed explicit notation of this dependence for clarity.

Of the three $\delta Q_n$, only $\delta Q_r$ is an independent dynamical variable, since it couples to an inductance (the effective ``mass" for a fictitious particle whose ``position" is $\delta Q_r$ \cite{devoret,likharevbloch,*hekking}), while $\delta Q_b$ and $\delta Q_c$ are deterministically related to $\delta Q_r$. To determine these relations, we hold $V_b$ and $\delta Q_r$ fixed and minimize $U$ with respect to $\delta Q_c$ and $\delta Q_b$, to obtain: $\delta Q_b=0$ and $\delta Q_r/C_r=\delta Q_c(C_r^{-1}+C_c^{-1}+C_q^{-1})$. Combining this with eqs.~\ref{eq:Etot}-\ref{eq:expand}, and re-expressing the result in terms of $Q_r$, we find:

\begin{equation}
U(Q_r)=E_q(Q_{q0})+\frac{(Q_r+C_cV_q)^2}{2C_r^\prime}\label{eq:stillU}
\end{equation}

\noindent where we have defined the quantities \cite{crprime}:

\begin{eqnarray}
C_r^\prime&\equiv&C_r+\frac{C_cC_q}{C_c+C_q}\nonumber\\
\omega_r&\equiv&\frac{1}{\sqrt{L_rC_r^\prime}},\;\;\;Y_r\equiv\sqrt{\frac{C_r^\prime}{L_r}}=\frac{1}{Z_r}\label{eq:omegar}
\end{eqnarray}

\noindent Thus, the leading-order effect of the coupling can be described as a displacement of the resonator mode's equilibrium ``position" by a qubit-state-dependent amount $C_cV_q$, as illustrated in fig.~\ref{fig:system}(c) and discussed in more detail in appendix~\ref{a:quasicharge}. This displacement can be associated with an effective ``force" $V_qC_c/C_r^\prime$ exerted by the qubit on the oscillator (as described by eq.~\ref{eq:Hpar} above), whose ``spring constant" is $1/C_r^\prime$. The next order effect arises from the qubit's dynamic capacitance $C_q$ \cite{LQmooij,*CQdelsing,*CQhakonen}, which is in general qubit-state-dependent, and can therefore induce small state-dependent shifts in the resonator's frequency and impedance according to eqs.~\ref{eq:omegar}; these shifts can be a potential source of gate errors, as we discuss in detail below in section~\ref{s:errors}.

\section{Controlled-phase gate with quasiclassical forces}\label{s:gates}

The form of eq. \ref{eq:stillU} and its interpretation in terms of a qubit state-dependent force on the resonator suggests an analogy with techniques developed for trapped atomic ions, in which state-dependent light forces acting on the atomic center-of-mass motion (analogous to the resonator in our case) produce entangling operations on the internal states of the atoms. We now show how similar entangling gates can be implemented in our system based on this analogy (discussed in more detail in appendix~\ref{a:ioncomp}). Following refs.~\onlinecite{MSPRL,*MSPRA,milburn,Znature,haljan}, we describe the oscillator in terms of a dimensionless classical field amplitude: $\alpha= \langle\hat u+i \hat v\rangle$, where:

\begin{eqnarray}
\hat u&\equiv&\frac{\hat Q_r}{\sqrt{\hbar Y_r}}=\frac{\hat a+\hat a^\dagger}{\sqrt{2}}\nonumber\\
\hat v&\equiv&\frac{\hat\Phi_r}{\sqrt{\hbar Z_r}}=\frac{\hat a-\hat a^\dagger}{i\sqrt{2}}\label{eq:uv}
\end{eqnarray}

\noindent and $\Phi_r$ is the canonical momentum of the oscillator such that $[\hat u,\hat v]=i$. The quantity $\alpha$ is a c-number for a pure coherent state, while in the presence of thermal or other classical field fluctuations it can be written as a diagonal density matrix (e.g., using the Glauber-Sudarshan $P$-representation \cite{Prep}). When we couple $N$ qubits to our resonator, in general all classical resonator quantities become operators in the $2^N$-dimensional qubit Hilbert space: $\alpha\rightarrow\hat\alpha$, $\omega_r\rightarrow\hat \omega_r$, $Z_r\rightarrow\hat Z_r$, $Y_r\rightarrow\hat Y_r$. In the cases of interest to us here, these operators are all \textit{diagonal}, and commute with the individual qubit Hamiltonians \cite{longit}.

We wish to focus on the qubit-state-dependent classical oscillator dynamics of the operator $\hat\alpha$, while the state dependences of $\hat\omega_r$ and $\hat{Z}_r$ implied by eq.~\ref{eq:omegar} are higher-order effects which we will consider as perturbations only, in section~\ref{s:errors}. To that end, we separate out these dependences explicitly with the notation:

\begin{eqnarray}
\hat\omega_r&\equiv&\tilde\omega_r(1+\delta\hat\omega_r)\nonumber\\
\hat{Z}_r&\equiv&\tilde{Z}_r(1+\delta\hat{Z}_r)\label{eq:tilde}
\end{eqnarray}

\noindent where the tilde is defined by: $\tilde{X}_r\equiv \textrm{Tr}[\rho_m\hat{X}_r]$, with $\rho_m\equiv\hat I_{2^N}/2^N$ the completely mixed state of the $N$ qubits, and $\hat{I}_d$ is the $d$-dimensional identity matrix. This definition separates the \textit{average} effect on the resonator due to coupling to the qubits (a renormalization of its frequency and impedance) from the small qubit-state-dependent corrections which we will consider as potential sources of error in section~\ref{s:errors}. We can now write the equation of motion for the oscillator with quality factor $Q$ as \cite{validity}:

\begin{equation}
\frac{d\hat\alpha}{d\tau}=-i(1+\delta\hat\omega_r)[\hat\alpha-\hat\eta]-\frac{\hat\alpha-\hat\alpha^*}{2Q}\label{eq:osc}
\end{equation}

\noindent where $\tau\equiv\tilde\omega_r t$, we have made the following replacements in eqs.~\ref{eq:uv} above: $Z_r=1/Y_r\rightarrow \tilde Z_r=1/\tilde Y_r$, and the dimensionless, qubit-state-dependent force $\hat\eta$ and frequency shift $\delta\hat\omega_r$ can be written:

\begin{eqnarray}
\hat\eta&\equiv&\sum_{i=1}^N\left[(\eta^-+\delta\eta^-)\hat\sigma_i^z+(\eta^++\delta\eta^+)\hat{I_i}\right]\nonumber\\
\delta\hat\omega_r&\equiv&\sum_{i=1}^N\delta\omega_r^-\hat\sigma_i^z\label{eq:eta_om_defs}
\end{eqnarray}

\noindent where $i\in1...N$ indexes the qubits, $\hat\sigma^z_i$ and $\hat I_i$ are the Pauli-$z$ and identity operator for qubit $i$, and:

\begin{eqnarray}
\eta^\pm&\approx&\frac{C_cV_q^\pm}{2\sqrt{\hbar \tilde Y_r}}\nonumber\\
\delta\omega_r^-&\approx&\frac{\beta_r\beta_q^-}{4}\nonumber\\
\beta_r&\equiv&\frac{C_c}{C_c+C_r}\;,\;\;\hat\beta_q\equiv\frac{C_c}{C_c+\hat C_q}\label{eq:eta_om_defs2}
\end{eqnarray}

\noindent using the notation: $X^\pm\equiv [\langle e|\hat{X}|e\rangle\pm\langle g|\hat{X}|g\rangle]$ with $\hat X$ a single-qubit operator. For the sake of clarity we have assumed all $N$ qubits and coupling elements are identical (though the method we propose does not require this), and we have retained only the leading-order term in $\delta\omega_r^-\ll 1$ (we will see in table~\ref{tab:sample} below that this is a very good approximation). Note that there is no nonzero $\delta\omega_r^+$ because of our definition of $\tilde\omega_r$ [c.f., eq.~\ref{eq:tilde}]. The quantities $\delta\eta^\pm$ are Langevin noise terms, with $\delta\eta^-$ due to qubit bias noise, and $\delta\eta^+$ associated with the finite resonator $Q$. In our model [fig.~\ref{fig:system}(a)] the latter comes from the Johnson-Nyquist noise $\delta V_r$ of the resistance $R_r$ ($Q=Z_r/R_r$). The resulting dimensionless noise power spectral density of the fluctuating force $\delta\eta^+$ can be written:


\begin{eqnarray}
S_{\delta\eta^+}(\Omega)&=&\frac{C_r^2\tilde\omega_r}{\hbar\tilde Y_r}\langle\delta V_r(t)\delta V_r(t^\prime)\rangle_\omega\nonumber\\
&=&\frac{2\Omega}{Q}\coth\left(\frac{\Omega\tau_c}{2}\right)\label{eq:fluct}
\end{eqnarray}

\noindent where the brackets denote an environment average, the subscript $\omega$ indicates a Fourier transform, $\Omega\equiv\omega/\tilde\omega_r$ is dimensionless frequency, and $\tau_c\equiv\hbar\tilde\omega_r/k_BT_r$ with $T_r$ the effective resonator mode temperature.

\begin{figure}
\includegraphics[width=3.25in]{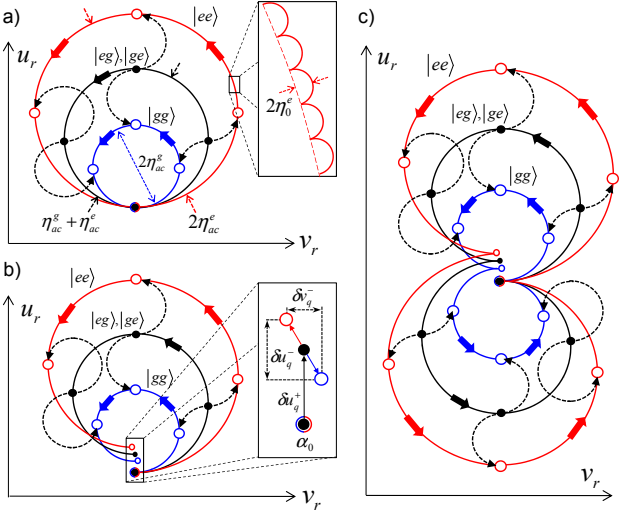}
\caption{\label{fig:phasespace} Oscillator phase space trajectories due to classical ``forces" exerted by two qubits. Trajectories are shown in a frame rotating at $\tilde\omega_r$, and the area enclosed within each path, indicated by solid lines ($|gg\rangle$ - blue, $|eg\rangle$,$|ge\rangle$ - black, $|ee\rangle$ - red) is the geometric phase acquired by that amplitude of the qubits' internal-state wavefunction. Dashed arrows show the displacements of $|ee\rangle$ and $|gg\rangle$ \textit{relative to} $|ge\rangle,|eg\rangle$, to make a connection between the general case considered here where $\langle eg|\hat\eta|eg\rangle,\langle ge|\hat\eta|ge\rangle\neq0$, and the usual situation in the trapped-ion case $\langle eg|\hat\eta|eg\rangle,\langle ge|\hat\eta|ge\rangle=0$ \cite{MSPRL,*MSPRA,haljan,Znature}. (a) In the ideal case ($Q\rightarrow\infty,\delta\hat\omega_r=0$), the paths are closed. The inset shows the high-frequency component of the response due to the counter-rotating term in eq.~\ref{eq:gateresult}. (b) for finite $Q,\delta\hat\omega_r$, the trajectories are no longer closed, such that at the end of the gate a nonzero total oscillator displacement $\delta\alpha_q^-$ is entangled with the qubits' internal state. The inset shows the components $\delta u_q^\pm,\delta v_q^-$ of this residual displacement ($\delta v_q^+=0$ because $\delta\omega_r^+=0$); finite decay results in a shrinking radius of the path with time, producing finite $\delta u_q^-$ at the end of the gate, and finite $\delta\hat\omega_r$ makes the modulation detuning $\delta_m$ weakly qubit-state-dependent, producing a nonzero $\delta v_q^-$. (c) modified gate sequence where either the force is reversed, or the qubits are inverted, in the middle of the gate, such that at the end of the gate the spurious entanglement between qubits and oscillator is removed.}
\end{figure}

To produce a gate which entangles a particular subset of the qubits coupled to a single resonator, we modulate the bias points of those qubits, as illustrated in fig.~\ref{fig:system}(c), while leaving the other (bystander) qubits alone. This modulation results in an oscillatory force on the resonator $\hat\eta=\hat\eta_0\sin(\omega_mt)$ which depends on the joint state of the qubits being modulated, and not on the state of the bystander qubits. Following the trapped-ion case \cite{Znature,MSPRL,*MSPRA}, we choose a modulation frequency $\omega_m$ close to resonance with a particular resonator mode, whose field (described by the classical, complex $\alpha$) begins to follow a qubit-state-dependent path in its $u,v$ phase space, as illustrated in fig.~\ref{fig:phasespace}. Each amplitude in the qubits' Hilbert space then begins to accrue a geometric phase associated with the phase-space area enclosed by the corresponding oscillator path: $\hat\phi_g=\textrm{Im}[\oint\hat\alpha^*d\hat\alpha]$ (recall that the operator notation here refers only the qubit Hilbert space; the resonator field is being treated as a classical, complex number in each dimension of this space). Neglecting for the moment the finite $Q$, fluctuations $\delta\eta^\pm$, and frequency shift $\delta\omega_r^-$, and taking the modulation to be turned on at $\tau=0$ with the oscillator in the state $\alpha_0$ and not entangled with the qubit(s), we find (in a frame rotating at $\Omega_m\equiv\omega_m/\tilde\omega_r$):

\begin{eqnarray}
\hat\alpha&=&\hat{I}_N\alpha_0e^{i\delta_m\tau}\label{eq:gateresult}\\
&\;\;&+\;i\hat\eta_{ac}\left[(\Omega_me^{i\delta_m\tau}-1)-\frac{\delta_m}{2}(e^{i2\Omega_m\tau}+1)\right]\nonumber\\
\hat\eta_{ac}&\equiv&\frac{\hat\eta_0}{\delta_m(2+\delta_m)}\nonumber
\end{eqnarray}

\noindent where the detuning $\delta_m$ of the modulation is defined according to: $\omega_m\equiv\tilde\omega_r(1+\delta_m)$ \cite{static}. In eq.~\ref{eq:gateresult}, the first term is free evolution, and the second and third terms can be viewed as the co- and counter-rotating components of the oscillator response, respectively, due to the effective force $\hat\eta_{ac}$. These oscillator dynamics in phase space, for a two-qubit system, are illustrated in fig.~\ref{fig:phasespace}(a).

Note that the effective force $\hat\eta_{ac}$ is inversely proportional to the detuning $\delta_m$ between the modulation frequency and the oscillator resonance, such that even if the resonator has additional internal modes, a single, desired mode can be selectively excited by tuning sufficiently close to resonance with it. We show in appendix~\ref{a:hm} that this selectivity is more than sufficient to completely neglect spurious excitation of other modes in practical cases. It also allows multiple modes of the resonator to be used intentionally, in parallel, if they are reasonably well-separated and the resonator is linear. This makes it possible in principle both to perform entangling operations simultaneously on multiple, distinct subsets of the qubits coupled to a single resonator, or to perform much faster operations on a single subset of those qubits.

Qubit state-dependent forces necessarily result in entanglement between the qubits' internal states and the resonator mode, as shown in fig.~\ref{fig:phasespace}. Although this type of entanglement has interesting applications in its own right, for the present purpose we only wish to operate on the qubits, using the resonator as a tool. Furthermore, entanglement with the resonator necessarily makes the qubits susceptible to additional decoherence. Therefore, the aim of our proposed operations (following the atomic case \cite{milburn,MSPRL,MSPRA}) is to remove this entanglement immediately, such that the net result is an operation only on the qubits. As illustrated in fig.~\ref{fig:phasespace}(a) for the ideal case, this occurs naturally because the phase space trajectories are approximately circular (essentially due to a beating between the modulation and the resonant component of the oscillator's response to it) such that at certain times they all return to the initial point, and the entanglement of the oscillator with the qubits vanishes. These times $\tau$ satisfy the conditions:

\begin{equation}
\frac{\delta_m\tau}{2\pi}=m,\;\;\;\;\frac{\tau}{2\pi}=\frac{k}{2} \label{eq:integers}
\end{equation}

\noindent where $m,k$ are integers with $k>2m$. Although the oscillator returns to its initial state, within the qubit subspace the system has accrued a state-dependent geometric phase, given by (neglecting overall phases):

\begin{equation}
\hat\phi_g=\frac{2\pi m}{\delta_m^2(2+\delta_m)} [(\eta_0^-)^2\hat\sigma_1^z\hat\sigma_2^z+2\eta_0^-\eta_0^+(\hat\sigma_1^z+\hat\sigma_2^z)]\label{eq:twoqb}
\end{equation}

\noindent A controlled-$\pi$ gate, sufficient in combination with single-qubit gates for universal quantum operations, is implemented (in addition to the single-qubit phase shifts given by the second term in eq.~\ref{eq:twoqb}, which one would need to correct using single-qubit rotations) if we choose:

\begin{equation}
\eta_0^-=\delta_m\sqrt{\frac{2+\delta_m}{4m}}\label{eq:eta0}
\end{equation}

\noindent in a total gate time $\tau_\pi$ (to leading order in $\delta_m^{-1}$) of:

\begin{equation}
\frac{\tau_\pi}{2\pi}\approx\frac{\delta_m}{2(\eta_0^-)^2}\label{eq:tf}
\end{equation}

\noindent when eqs.~\ref{eq:integers} and~\ref{eq:eta0} are satisfied.

Before discussing realistic errors in these operations in the next section, we highlight the qualitative distinction between the type of coupling we have been discussing and that used in cQED \cite{blaisarch,*schusterCQED,wallraffCQED,*noriCQED}. This can be compactly expressed by writing the effective coupling strength in both cases as the time-derivative of the two-qubit conditional phase (for our gates, the first term of eq.~\ref{eq:twoqb}):


\begin{equation}
\hbar\dfrac{d\phi_c}{dt}\approx
\begin{cases}
\dfrac{\beta_r^2}{2\delta_m}\times\dfrac{1}{2}C_r(V_q^-)^2, & \text{this work} \\[3ex]
\dfrac{\beta_r^2}{2\delta_q}\times\hbar\omega_r\dfrac{Z_r}{R_Q}, & \text{cQED}
\end{cases}
\end{equation}

\noindent where in the latter case $\delta_q\equiv\Delta_q/\omega_r$ is the dimensionless qubit-resonator detuning, and $R_Q\equiv h/4e^2$ is the superconducting resistance quantum. Notice that for our gate the energy scale of the effective interaction is purely classical and completely independent of the qubit frequency $\omega_q$, since it is based only on deterministic, classical, driven resonator dynamics. By contrast, the cQED effective interaction energy is explicitly quantum, with the factor $Z_r/R_Q$ describing zero-point fluctuations of the resonator ground state (inductive coupling in cQED would give the inverse of this factor, $R_Q/Z_r$), and it depends explicitly on the detuning between qubit and resonator $\Delta_q$. These features are a natural consequence of virtual photon exchange between qubit and resonator driven by vacuum fluctuations.

Because of this strongly quantum nature of the interaction in cQED, any classical fluctuations of the resonator must be negligible compared to its zero-point quantum fluctuations if high-fidelity operations are to be performed. As described in ref.~\onlinecite{sears}, the passage of a single spurious photon through any strongly-coupled resonator mode during a quantum operation effectively makes a projective measurement of all the qubits coupled to it. Such fluctuations of the resonator are the inevitable result of photon exchange with transmission lines \cite{sears,rigetti} and qubits \cite{esteve,*corcoles} to which it is coupled, which often have a significantly higher effective temperature than the bath to which the experiment is nominally anchored. Thus, extremely careful filtering over a wide frequency range is necessary to suppress these fluctuation-induced errors in cQED \cite{sears}. By contrast, we will show that for our gates resonator fluctuations have only a higher-order effect, and only the specific mode near the modulation frequency contributes to this effect.

\section{Two-qubit gate fidelity} \label{s:errors}

In this section we evaluate sources of error in our two-qubit controlled phase gate. The results obtained in this section are tabulated in table~\ref{tab:sample} for a number of different physical qubit modalities and parameter assumptions. We begin with the effect illustrated in fig.~\ref{fig:phasespace}(b), spurious entanglement between qubits and resonator at the end of the gate in the form of a residual qubit-state-dependent oscillator displacement $\delta\alpha_q^-$. The two mechanisms which produce this type of entanglement are finite resonator $Q$, and the state-dependent frequency shift $\delta\hat\omega_r$ (induced by a state-dependent $C_q$ or $L_q$). As shown in fig.~\ref{fig:phasespace}(b), resonator decay causes the radius of the phase-space trajectory to shrink with time, while $\delta\hat\omega_r$ effectively produces a detuning $\delta_m$ (and corresponding evolution rate around the paths in phase space) which is qubit-state-dependent. These result in the spurious displacements illustrated in the inset: $\delta u_q^-\sim\mathcal{O}(1/Q)$ and $\delta v_q^-\sim\mathcal{O}(\delta\omega_r^-)$, respectively.

Since both of these displacements result from classical and deterministic dynamics, however, we can strongly suppress them using the scheme shown in fig. \ref{fig:phasespace}(c). We divide the gate excitation into two periods of equal duration (\textit{each} of which satisfies eqs. \ref{eq:integers}), and we switch the sign of the effective force $\eta_{ac}$ between these two periods (alternatively, one could insert $\pi$-pulses to invert the qubits instead). This simple procedure works like a spin-echo, in the sense that the classical modifications to the trajectories cancel out at the end of the operation, removing the leading terms of order $\sim\mathcal{O}(1/Q),\mathcal{O}(\delta\omega_r^-)$ in the final displacement. The resulting gate error is, to leading order in the small quantities $\delta_m,\eta_0^-,Q^{-1},\delta\omega_r^-$, and $\bar{n}$ (see appendix~\ref{a:errors}):

\begin{equation}
\epsilon_{\delta\alpha}\approx \frac{mx^4}{4}\left[1+2\bar{n}+2m\left(\frac{2\pi\delta\omega_r^-}{x}\right)^2\right]\label{eq:epsdalpha}
\end{equation}

\noindent where $x\equiv\pi/(Q\eta_0^-)$ and $m$ includes both halves of the gate. As we show in table~\ref{tab:sample}, the result is essentially negligible compared to other error sources for all of the cases considered.

The more important potential source of errors is fluctuations of the qubits or resonator, which can produce single-qubit dephasing and fluctuations of the controlled-phase imparted during the entangling operation. One example of this is noise local to the qubits themselves, such as the ubiquitous 1/f charge and flux noise encountered in superconducting circuits \cite{ithier}. Such noise is particularly important for qubits that do not have a degeneracy point like that shown by point A in fig.~\ref{fig:system}(c) (where $V_q^-=0$ or $I_q^-=0$) and are therefore sensitive to low noise frequencies (for example, singlet-triplet quantum dots \cite{petta,*laird,spinQDs2,QDcouple}). Dynamical decoupling techniques have been extensively developed to suppress this sensitivity \cite{biercuk,*bylander}, the simplest example of which is the spin-echo \cite{ithier}. In our gate of fig.~\ref{fig:phasespace}(c) this could be naturally implemented by replacing the reversal of the gate force $\eta_{ac}$ in the middle of the controlled-$\pi$ gate with a $\pi$-pulse on each qubit, exchanging the roles of $|g\rangle$ and $|e\rangle$ (similar to ref.~\onlinecite{QDcouple}). Since these errors are entirely independent of our proposed method, we will not discuss them further here.

Higher-frequency noise (i.e. fluctuations \textit{during} the gates), however, is an important source of error which we now consider. Assuming small fluctuations $\delta\eta^\pm$ of the force about the desired values, the system will accrue an additional geometric phase:

\begin{eqnarray}
\delta\phi_g\approx2\int\biggl\{\delta\eta^-\bigl[\eta^+(\hat\sigma_1^z+\hat\sigma_2^z)+\eta^-\hat\sigma_1^z\hat\sigma_2^z\bigr]\nonumber\\
+\;\delta\eta^+\bigl[\eta^-(\hat\sigma_1^z+\hat\sigma_2^z)\bigr]\biggr\}d\tau\label{eq:deltaphig}
\end{eqnarray}

\noindent where the three terms are: single- and two-qubit phase errors due to fluctuations of the state-dependent force (predominantly qubit bias noise), and single-qubit errors due to oscillator fluctuations (qubit bias noise can also contribute to this if $V_q^+\neq0$ or $I_q^+\neq0$). The resulting contributions to the average error can then be expressed in terms of mean square phase fluctuation amplitudes of the form:

\begin{equation}
\langle\delta\phi_g^2\rangle\sim\int S_{\delta\eta^+}(\Omega) |\eta^-(\Omega)|^2d\Omega\label{eq:deltaphig2}
\end{equation}

\noindent where $S_{\delta\eta^+}(\Omega)$ is a dimensionless noise power spectral density (of the fluctuations $\delta\eta^+$ corresponding to the third term in eq.~\ref{eq:deltaphig}), $\eta^-(\Omega)$ is the fourier transform of the time-dependent gate force $\eta^-(\tau)$, which is a peaked function centered on $\Omega_m$, of width $\sim(2\tau_\pi)^{-1}$ and amplitude $\sim\eta_0^-\tau_\pi$. Since the gate forces $\eta^\pm$ oscillate at the frequency $\omega_m$, only noise which occurs at or near this high frequency (or to a lesser extent its harmonics) will produce errors \cite{biercuk,*bylander}. Because of this, in nearly all cases low-frequency qubit bias noise (e.g. 1/f charge or flux noise) can be ignored for our gates, allowing us to neglect the contributions of the fluctuations $\delta\eta^-$ associated with the first two terms in eq.~\ref{eq:deltaphig}.

The dominant source of high frequency noise in our model is then the thermal oscillator noise, which appears in $\delta\eta^+$, and which can become quite important as the $Q$ of the oscillator is reduced. Using eqs.~\ref{eq:fluct} and~\ref{eq:deltaphig2}, we obtain the average error (per qubit):

\begin{equation}
\epsilon_{\delta\eta^+}\sim\frac{\pi}{\sqrt{2m}Q\eta_0^-}\coth{\frac{\tau_c}{2}}\label{eq:Qfluct}
\end{equation}

\noindent These errors will tend to restrict how small the resonator mode $Q$ can be \cite{T1}. Note, however, that fluctuations of other resonator modes can be neglected; as described in appendix~\ref{a:hm}, even the driven excursions of higher modes are negligible (at the error rates of interest here) due to the spectroscopic selectivity associated with driving the system near a specific, chosen mode.

\begin{center}
\begin{table*}
\begin{threeparttable}
    \begin{tabular}{  | c | cccccccccccccc |}
    \hline\noalign{\smallskip}
    qubit & $Q$ & $\omega_r/2\pi$ &  $L_r$ & $C_r$ &  $\eta_0^-$ & $\delta_m$ &  $m$ & $\Gamma_\phi^{-1}$ & $t_\pi$  & $\epsilon_{\delta\alpha}$ & $\epsilon_{\delta\eta^+}$ & $\epsilon_{2qb}$ & $\Delta_{hm}$ & $N_\gamma$\\
    type &  & [GHz] & [pH] & [pF] & $[\times10^3]$  &$[\times10^3]$  &  & [ms] & [ns]   & $[\times10^3]$ & $[\times10^3]$ & $[\times10^3]$ & [GHz] &\\
    \hline
    \hline\noalign{\smallskip}
    Flux& $25000$ & 10 &  250 & 0.84 &  4.5 &  64  & 100 &8.4& 160&   0.015 & 2.0 & 2.0 & 2.3 & 2600\\
    \hline\noalign{\smallskip}
    \cite{mooij,*MITtunable}&$10^6$& 10 &  500 & 0.46 &  3.4 &  12  & 6 & 0.76 & 51  & 10$^{-9}$ & 0.27 & 0.34 & 3.5 & 5200 \\ 
    \hline\noalign{\smallskip}
    &$10^6$& 1 &  30nH & 0.84 &  1.4 &  2.9  & 2 & 0.23 & 700 & $2\times 10^{-8}$ & 2.0 & 5.0 & 0.03 & $3.1\times10^6$ \\
    \hline\noalign{\smallskip}
    Transmon& $50000$ & 10 &  250 & 0.84 &  1.4 &  14  & 50 & 126 & 360 & 0.053 & 4.5 & 4.6 & 0.4 & $10^4$ \\
    \hline\noalign{\smallskip}
    \cite{houck}& $10^6$ & 10 &  250 & 0.84 &  1.4 &  6.2  & 10 & 6.3 & 160 & $7\times 10^{-8}$ & 0.51 & 0.53 & 0.9 & $10^4$\\
    \hline\noalign{\smallskip}
    & $10^6$ & 3 &  1500 & 1.8 & 1.1 &  2.2  & 2 & 0.10 & 300 & $3\times 10^{-8}$ & 1.5 & 4.5 & 0.07 & $2.0\times10^5$\\
    \hline\noalign{\smallskip}
    S-T QDs& $25000$ & 15 &  100nH\footnotemark[1] & 1.1fF & 3.0 & 33  & 60 & 3.0 & 120 & 0.20 & 3.8 & 4.0 & 1.4 & 2.5 \\
    \hline\noalign{\smallskip}
    \cite{petta,*laird,spinQDs2,QDcouple}& $10^6$ & 15 &  30nH\tnote{a} & 3.7fF & 1.7 &  6.6  & 8 & 0.83 & 80 & $5\times 10^{-7}$ & 0.47 & 0.57 & 1.4 & 8.5\\
    \hline\noalign{\smallskip}
    CaCl$^+$\cite{molions}& $10^6$ & 1 &  1000nH\tnote{a} & 25fF &  0.14 &  1.1  & 30 & 25s & 27$\mu$s  & 0.0032 & 5.1 & 5.1 & 0.01 & $2.4\times10^9$\\
    \hline\noalign{\smallskip}
    $^{31}$P-$^{28}$Si \cite{Si}& $10^7$ & 1 &  100\tnote{b} & 250 &  0.014 &  0.14  & 50  & $2\times10^{7}$s & 350$\mu$s & 0.005 & 4.0 & 4.0 & 0.5 MHz & $2.4\times10^9$\\
    \hline\noalign{\smallskip}
    NV \cite{balaNV}& $10^7$ & 1 &  100\tnote{b} & 250 &  0.028 &  0.13  & 10  & 2s & 78$\mu$s & $6\times10^{-5}$ & 4.4 & 4.5 & 0.8 MHz & $2.4\times10^9$ \\
    \hline\noalign{\smallskip}
   \end{tabular}

   \begin{tablenotes}
   \item[a]\footnotesize Impedances this high require the use of high-kinetic-inductance materials \cite{kermanKI,*barendsKI,*kamlapureKI,*baekKI,*sandbergKI,MRFS}. Limits on achievable resonator $Q$ and impedance for these materials are as yet unknown.
   \item[b]\footnotesize Impedances this low may not be achievable in a transmission-line resonator.
   \end{tablenotes}

      \caption{\label{tab:sample} Selected examples of two-qubit controlled-$\pi$ gate parameters (details for each system are contained in appendix~\ref{a:modalities}). For transmission-line resonators, $L_r$ and $C_r$ are effective values for a given longitudinal mode. The results for gate errors shown are the thermal photon-number dephasing rate $\Gamma_\phi$ [eq.~\ref{eq:nthdephase}] (which is present due to the coupling even when no two-qubit gates are being driven), the gate time $t_\pi$ [eq.~\ref{eq:tf}], the state-dependent displacement error $\epsilon_{\delta\alpha}$ [eq.~\ref{eq:epsdalpha}] and the resonator fluctuation error $\epsilon_{\delta\eta^+}$ [eq. \ref{eq:Qfluct}]. The total two-qubit error is defined as: $\epsilon_{2qb}\equiv2\Gamma_\phi t_\pi+\epsilon_{\delta\alpha}+\epsilon_{\delta\eta^+}$. This does not include single-qubit errors unrelated to the coupling (for example, $T_1$ relaxation for the superconducting qubits or charge-noise dephasing for the quantum dots). The quantity $\Delta_{hm}$ is the minimum detuning to the next higher oscillator mode (assuming it has the same Lamb-Dicke parameter as the fundamental), such that the associated error $\epsilon_{hm}\le0.1\times\epsilon_{2qb}$ [eq.~\ref{eq:ehm}]. In all cases we take a resonator temperature of $T_r=40$ mK, corresponding to $\bar{n}=0.4,0.03,4\times10^{-6},10^{-8}$ for $\omega_r/2\pi=1,3,10,15$ GHz. Note that both donor spins in Si and NV centers in diamond do not themselves require low temperatures, but here they are required to suppress errors due to classical, thermal resonator fluctuations [c.f., eqs.~\ref{eq:Qfluct},~\ref{eq:nthdephase}]).}

   \end{threeparttable}

   \end{table*}
\end{center}


In addition to the geometric phase errors just discussed in the controlled-phase gate, thermal resonator fluctuations can also produce direct dephasing of the qubits even when there is no gate modulation, if $\delta\omega_r^-$ is nonzero. This occurs because a qubit-state-dependent frequency shift of the resonator can also be viewed as an effective qubit frequency splitting $\omega_{qb}^\prime$ that depends on the resonator photon number $n$: $\omega_{qb}^\prime=\omega_{qb}+2n\delta\omega_r^-$ \cite{blaisarch,*schusterCQED}. For a thermal photon number distribution at temperature $T_r$ with mean photon number $\bar{n}$ \cite{cohn}, each qubit then experiences dephasing at the rate \cite{blaisarch,*schusterCQED}:

\begin{equation}
\Gamma_\phi\approx 16\tilde\omega_r\bar{n} Q(\delta\omega_r^-)^2\label{eq:nthdephase}
\end{equation}

\noindent Notice that this dephasing rate \textit{increases} with increasing resonator $Q$, as the discrete resonator frequencies associated with different photon number states become more resolved \cite{blaisarch,*schusterCQED}. This will tend to restrict how large the resonator $Q$ can be.

Using eqs.~\ref{eq:epsdalpha}, \ref{eq:Qfluct}, and \ref{eq:nthdephase}, we list in table ~\ref{tab:sample} the parameter values and resulting gate errors for a number of specific examples, chosen to illustrate the utility of our scheme over a range of physical qubit modalities, resonator $Q$s, resonance frequencies, and thermal photon populations $\bar{n}$. Note that in some cases two-qubit error rates as low as $10^{-3}$ are still achievable with $\bar{n}$ as high as 0.4 (for a 1 GHz resonator at 40 mK) and $Q$ as low as 25,000, showing the robustness of our technique against resonator fluctuations and decay. For state-of-the-art experimental values such as: $Q\sim10^6$ \cite{martinisAPL} and $\bar{n}\sim10^{-3}$ \cite{rigetti}, even lower error rates $\sim10^{-4}$ become accessible (provided of course that single-qubit errors unrelated to the operations considered here are not the limiting factor).

We note in this context that while very high resonator $Q$s of $10^6$ or greater are at present difficult to achieve in planar geometries at the single-photon level required for cQED implementations, this difficulty is substantially reduced at higher resonator powers ($\sim10^3-10^4$ photons) where parasitic, lossy defects become saturated \cite{martinisAPL}. Given that our proposed method is in principle not affected by a coherent initial resonator state, one might consider the prospect of intentionally driving the resonator to increase the effective $Q$ for gate operations (note, however, that the thermal fluctuation errors discussed above associated with $T_r$ would also result from a nonzero effective temperature of an additional drive field). The simplest way to accomplish this would be to drive the resonator \textit{instead} of the qubits, and use the qubit/resonator coupling to accomplish the modulation of $\hat\eta$ for the gate. In table \ref{tab:sample} we list the quantity $N_\gamma$, the resonator photon number required to produce by itself the full qubit bias swing (via its coupling to the qubits) assumed for each set of gate parameters. In most cases, these values are comparable to or larger than than the $\sim10^3-10^4$ photons typically necessary to saturate the resonator loss in current experiments \cite{resdrive}.

\section{Conclusion}\label{s:conclusion}

We have described an alternative method for coupling qubits and resonators which is qualitatively distinct from the current circuit QED paradigm. Unlike cQED, in which real or virtual photon exchange between qubits and resonator mediates the interaction \cite{blaisarch,*schusterCQED,wallraffCQED,*noriCQED}, our proposal is based on a first-order, \textit{longitudinal} coupling which does not involve any photon exchange, and which relies on purely classical dynamics of the resonator. We have shown how this coupling can be understood as a qubit-state-dependent effective force acting on the resonator, in a manner analogous to that which has been engineered between the internal spin states and center-of-mass vibrational modes of trapped atomic ions \cite{MSPRL,*MSPRA,milburn,Znature,haljan}. Our method has some potentially advantageous features when compared with cQED: first, since no photons are exchanged between qubits and resonator, there is no Purcell effect \cite{houck} and the qubits' excited-state lifetimes are decoupled both from the qubit-resonator detuning, and from the resonator $Q$. In fact, our method does not require the qubits to have any nonzero transition dipole moment at all, which opens the possibility of qubit designs that are intrinsically decoupled and therefore potentially much longer-lived \cite{MRFS}. The lack of photon exchange also implies that the coupling is independent of qubit frequency, which allows all qubits that have a degeneracy point to be biased at that point, so that they can remain insensitive to low-frequency noise even during quantum operations. Limitations associated with using the detuning between qubits and cavities to control the couplings \cite{helmer,*rezQu,*divincenzo} are largely removed, in particular with regard to on/off coupling ratios and to implementation of complex, highly-interconnected, many-qubit systems. Next, unlike in cQED where gate operations are linearly sensitive to the presence of spurious photons in any resonator mode coupled to the qubits \cite{blaisarch,*schusterCQED,sears,rigetti,multimode}, our scheme is only sensitive to occupation of a single, spectroscopically selected resonator mode, in higher order, such that it can tolerate substantial thermal occupation of that mode (up to $\bar{n}=0.4$ was considered in table~\ref{tab:sample}) before significant errors occur. This mitigates the need for an extremely low effective resonator temperature and/or a high resonator frequency (i.e., $T_r\ll\hbar\omega_r/k_B$). Also, since a classical drive field in the resonator does not produce errors in our gates, it may even be possible to intentionally drive it into the regime where high $Q$ is much easier to achieve than in the single-photon limit required for cQED \cite{martinisAPL}. Third, because our coupling scheme gives an effective interaction $\propto(\eta\sum_i\hat\sigma_i^z)^2$ \cite{MSPRL,*MSPRA}, its strength does not decrease as the number of interacting qubits increases. By contrast, in cQED, multiqubit interactions must either be engineered by cascading pairwise interactions \cite{neeley,*dicarlo,martinisALG,*reedALG} or using weaker, higher-order multiqubit couplings involving more than one virtual exchange of photons with the resonator. Multiqubit interactions may be of interest, for example, in cluster-state generation \cite{cluster}, or syndrome extraction in quantum error-correction schemes such as surface codes \cite{surface} and low-density parity check codes \cite{pryadko}. Finally, our scheme can also be used as a QND readout technique: one simply modulates the qubit bias at the oscillator resonance, such that the field amplitude and/or phase to which the oscillator rings up depends on the state of the qubit. Such a readout has the important advantage that it does not suffer from the dressed dephasing effects encountered in conventional dispersive readout in cQED \cite{dressed}.

\textit{Note added} - An error-suppression method closely related to that shown in fig.~\ref{fig:phasespace}(c) was recently demonstrated for trapped atomic ions in [D. Hayes, S.M. Clark, S. Debnath, D. Hucul, I.V. Inlek, K.W. Lee, Q. Quraishi, and C. Monroe. Phys. Rev. Lett. {\bf 109}, 020503 (2012)].

We acknowledge helpful discussions with and/or comments from Daniel Greenbaum, Archana Kamal, Arthur Kerman, Adrian Lupa\c{s}cu, and William Oliver. This work is sponsored by the Assistant Secretary of Defense for Research \& Engineering under Air Force Contract \#FA8721-05-C-0002.  Opinions, interpretations, conclusions and recommendations are those of the author and are not necessarily endorsed by the United States Government.


\appendix

\section{Quasistatic resonator displacements}\label{a:quasicharge}

At the end of section~\ref{s:system}, we reached the conclusion that longitudinal coupling between the qubit and resonator as shown in fig.~\ref{fig:system}(a) produces a quasistatic displacement of the resonator's canonical position variable $Q_r$ [fig.~\ref{fig:system}(a)]. This may seem counterintuitive, given that there can be no quasistatic charge on the capacitor plates of $C_r$ due to the presence of $L_r$ across it. There is in fact no contradiction here, however, because the canonical position variable $Q_r$ is not simply the charge on the capacitor $C_r$. Rather, $Q_r$ is more precisely a \textit{quasicharge} \cite{likharevbloch,*hekking,AJKdual}, defined in this case by: $dQ_r/dt=I_r+C_rdV_r/dt$, whose corresponding canonical momentum $\Phi_r$ satisfies: $d\Phi_r/dt=V_r-L_rdI_r/dt$. A nonzero quasistatic displacement of $Q_r$ is therefore possible due to a nonzero time-integral of the displacement current $C_rdV_r/dt$.

To see intuitively how this arises, we can map our problem onto a more familiar electrostatic system, as shown in fig.~\ref{fig:familiar}(a): that of an electric displacement $C_cV_q$ contained inside the capacitor $C_r$ (where $V_q\equiv dE_q/dQ_q$). We can imagine this as an electric dipole which was moved into the capacitor (with its orientation held fixed), such that the resulting transient voltage appearing across $C_r$ gives $\int C_r(dV_r/dt)dt=C_c V_q$. This mechanical analogy casts the work done by the source (the second line of eq. \ref{eq:Etot}) into the form of a potential energy associated with ``assembling" the system, as is often encountered in electro- or magnetostatic problems: as a permanent electric dipole is moved into the resonator capacitor, there will be a mechanical force on that dipole due to the gradient of potential energy in real space, such that work must be done while moving the dipole into position. In our system, this ``motion" corresponds to turning up the bias voltage $V_b$.

Completely dual arguments to those just given apply to the magnetic case shown in fig. \ref{fig:system}(b), where the quasistatic displacement of $\Phi_r$ is given by $L_cI_q$, and $I_q\equiv dE_q/d\Phi_q$. The resonator capacitor $C_r$ prevents a quasistatic current from flowing in the inductor $L_r$, but since $\Phi$ is not simply the inductor flux but a ``quasiflux", it can be displaced quasistatically if an effective magnetization is moved into the inductor as illustrated in fig.~\ref{fig:familiar}(b). The resulting displacement is then the time integral of the induced emf $-L_rdI_r/dt$.

\section{Spurious state-dependent resonator displacements}\label{a:errors}

We seek to estimate the gate error that results when a nonzero entanglement between qubit state and oscillator displacement $\delta\alpha_q^-$ remains at the end of the gate. We can bound this error using the fidelity \cite{nielsen} between the desired qubit density matrix after the gate $\rho_f^{qb}$ and the trace over resonator states of the actual total density matrix after the gate $\rho_f^{tot}$ \cite{pointer}:

\begin{eqnarray}
\epsilon_{\delta\alpha}&\approx& 1-\left(\textrm{Tr}\left[\sqrt{(\rho_f^{qb})^{1/2}\textrm{Tr}_r[\rho_f^{tot}](\rho_f^{qb})^{1/2}}\right]\right)^2\\
\rho_f^{tot}&\equiv& D_Q^\dagger\left[\rho_f^{qb}\otimes D^\dagger(\alpha_0+\delta\alpha_q^+)\rho_{th}^rD(\alpha_0+\delta\alpha_q^+)\right]D_Q\nonumber\\
D_Q&\equiv& D\left[+\frac{\delta\alpha_q^-}{2}\right]|gg\rangle\langle gg|+D\left[-\frac{\delta\alpha_q^-}{2}\right]|ee\rangle\langle ee|\nonumber
\end{eqnarray}

\noindent where $\rho_{th}^r$ is a thermal resonator state with temperature $T_r$, $D(\delta\alpha)$ is the displacement operator in $u,v$ phase space for (complex) displacement $\delta\alpha$, and $D_Q$ performs the state dependent displacement shown in fig.~\ref{fig:phasespace}(b). As an approximate worst case estimate, we take: $\rho_f^{qb}=(|gg\rangle+|ee\rangle)(\langle gg|+\langle ee|)/2$, and find (using the results of ref.~\onlinecite{fidelity}):

\begin{figure}
\includegraphics[width=3.4in]{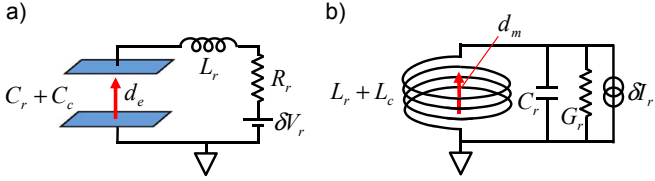}
\caption{\label{fig:familiar} Panels (a) and (b) show classical systems analogous to the circuits of figs.~\ref{fig:system}(a) and (b), in which an electric or magnetic displacement is placed inside the resonator, producing an orientation-dependent displacement of the resonator position.}
\end{figure}

\begin{equation}
\epsilon_{\delta\alpha}\approx\frac{|\delta\alpha_q^-|^2}{4}(1+2\bar{n})\label{eq:dalpha}
\end{equation}

\noindent to leading order in $\delta\alpha_q^-$ and $\bar{n}\approx\exp{(-\hbar\omega_r/k_BT_r)}$, the mean thermal photon number in the resonator, and independent of $\alpha_0+\delta\alpha_q^+$.

To evaluate the displacement $\delta\alpha_q^-$ due to nonzero resonator damping and finite $\delta\omega_r^-$, we first renormalize the oscillator resonance frequency to account for the usual $Q$-induced shift:

\begin{equation}
\tilde\omega_r\rightarrow\tilde\omega_r\sqrt{1-\frac{1}{4Q^2}}\label{eq:omegaro}
\end{equation}

\noindent and correspondingly renormalize the dimensionless time $\tau$ [eq.~\ref{eq:osc}], modulation frequency $\omega_m$, detuning $\delta_m$, and the conditions for $m$ and $k$ [eq.~\ref{eq:integers}]. The result, obtained by solving eq.~\ref{eq:osc} analytically and expanding to leading order in the small quantities $\delta_m,\eta_0^-,Q^{-1},\delta\omega_r^-$, is:

\begin{equation}
|\delta\alpha_q^-|^2\approx mx^4\left[1+2m\left(\frac{2\pi\delta\omega_r^-}{x}\right)^2\right]\label{eq:dalpha2}
\end{equation}

\noindent where $x\equiv\pi/(Q\eta_0^-)$ and $m$ now includes both halves of the gate. Combination of eqs.~\ref{eq:dalpha} and~\ref{eq:dalpha2} yields eq.~\ref{eq:epsdalpha}.

\section{Spurious excitation of higher resonator modes}\label{a:hm}

In the presence of the gate modulation force, higher modes in the resonator \cite{houck,multimode} will necessarily also be displaced by interaction with the qubits, according to their effective capacitance (inductance for the magnetic case of fig.~\ref{fig:system}(b)) and geometrical coupling factors. The resulting (state-dependent) excursions of these modes do not in general decouple from the qubit at the same time as the fundamental mode [c.f., fig.~\ref{fig:phasespace}], leaving a spurious entanglement between the qubits and the higher modes at the end of the gate. However, just as in the atomic case \cite{MSPRL,*MSPRA,milburn}, the excursions of higher modes are suppressed by their detuning from the modulation, since the effective force in the rotating frame is inversely proportional to that detuning [c.f., eq.~\ref{eq:gateresult}]. Using eq.~\ref{eq:dalpha} for the error due to a state dependent excursion $\delta\alpha_k^-$ of mode $k$ with frequency $\omega_{r,k}\equiv f_k\omega_r$, and taking $\delta\alpha_k^-$ to be $\sim\eta_{ac,k}^-$ the effective Lamb-Dicke parameter for that mode [c.f., eq.~\ref{eq:gateresult}], we obtain the following estimate for the error due to excitation of higher modes:

\begin{equation}
\epsilon_{hm}\sim\frac{(\eta_{0,k}^-)^2(1+2\bar{n})}{2\left(1-f_k^{-2}\right)^2}\label{eq:ehm}
\end{equation}

\noindent Using this result, we list in table ~\ref{tab:sample} the minimum separation $\Delta_{hm}$ from the fundamental to the next higher mode, assuming that $\eta_{0,k}^-=\eta_0^-$, which would give rise to an error $\epsilon_{hm}$ \textit{one tenth} of the total error from all other sources. In all cases $\Delta_{hm}<\omega_r$, indicating these errors are unlikely to be significant in practical cases.

One interesting possibility to consider in this context is that if the higher modes of the resonator are all commensurate with the fundamental (as is the case in an ideal transmission-line resonator), all of the higher modes will decouple from the qubits \textit{at the same time} as the fundamental. Since the geometric phases of all modes add together, this means that one could in principle implement a much faster gate by exciting many modes simultaneously using a modulation waveform which contains higher harmonics. The complication with this idea for real systems is that the higher modes are never quite commensurate with the fundamental, for example due to the reactance of the input coupling elements. To make this work, one would then need an appropriately tailored modulation waveform which selectively excites only those higher modes that are close enough to commensurate with the fundamental to keep the resulting errors low.

\section{Details on error estimates in table~\ref{tab:sample}}\label{a:modalities}

\subsection{Superconducting qubits}

For the transmon qubit [fig.~\ref{fig:examples}(a)], we take $\omega_{qb}(\Phi_{b0}=0)/2\pi=15$ GHz, which gives $L_q^-=L_q^+/2=-2(\Phi_0/\pi)^2/\hbar\omega_p=-87.4$ nH. We assume a modulation excursion $\delta\Phi_{AC}=0.2\Phi_0$, corresponding to $\omega_{qb}(0.2\Phi_0)=13.5$ GHz. For the flux qubit [fig.~\ref{fig:examples}(b)], we take $E_J/h=200$ GHz, $E_C/h=5.7$ GHz, and $\alpha(\Phi_{b0})=0.74$. This gives $\omega_{qb}(\Phi_{b0})/2\pi=4.73$ GHz and $L_q^-=15.9$ nH, $L_q^+=0$ \cite{qubitcalc}. We assume a modulation amplitude of $\delta\Phi_{AC}=0.1\Phi_0$, corresponding to $\alpha(0.1\Phi_0)=0.70$ and $\omega_{qb}(0.1\Phi_0)=7.8$ GHz. For both flux and transmon qubits we take $L_c=25$ pH \cite{circulate}. Also for both of these cases we must consider the effect of junction asymmetry in the DC SQUID (defined by the area asymmetry parameter: $A_J\equiv2(A_1-A_2)/(A_1+A_2)$), which produces a spurious coupling between external flux and the SQUID plasma mode. For the limit $L_c\ll L_J$ satisfied here, we find the matrix element for the modulation to couple to the first excited vibrational state of the plasma mode: $M_A\approx(\delta\Phi_{AC}/\Phi_0)A_J\sqrt{\pi/4}\hbar\omega_p$. We then have a maximal probability in this excited state of $\sim (\pi/4)(\delta\Phi_{AC}/\Phi_0)^2A_J^2(\omega_p/(\omega_p-\omega_m))^2$. Thus, for $A_J<0.05$ and $\delta\Phi_{AC}/\Phi_0<0.2$, this error is at the $10^{-4}$ level or below. This calculation also gives a residual, direct Jaynes-Cummings type coupling between DC SQUID plasma mode and the resonator: $g\approx\omega_p\beta_rA_J\sqrt{Z_r/8R_Q}$, which can be neglected in the cases considered here.

\subsection{Singlet-triplet coupled quantum dots}

Figure~\ref{fig:examples}(c): The singlet-triplet coupled quantum dots (and molecular ions below) are biased with a voltage, rather than a charge, so that: $C_b,C_c\gg C_q$ (the opposite limit from the superconducting qubit cases). In this limit, the voltages across $C_b,C_c$ can be neglected, and $V_q\approx V_b$. The quasicharge is then: $Q_q\approx dE^-/dV_q$ evaluated at $V_b$. The displacement of the resonator in eq.~\ref{eq:stillU} is then $C_qV_b$ and instead of eqs. ~\ref{eq:expand} and ~\ref{eq:eta_om_defs}, we have: $\eta_0^-\equiv (dE^-/dV_q)/(2\sqrt{\hbar Y_r})$, $C_q^-\equiv d^2E^-/dV_q^2$, and $\delta\omega_r^-/\omega_r\approx-C_q^-/(2C_r)$. For the quantum dots we use the parameters of ref.~\onlinecite{petta,*laird}, with a tunneling amplitude of $t_c=23\mu$eV. We take an exchange energy $J$ for each qubit which oscillates from $\sim 1\mu$eV to $\sim 2.5\mu$eV, corresponding to electrode voltages from -0.7mV to -0.4 mV, $dE^-/dV_e$ from 0.002$e$ to 0.01$e$, and $C_q^-=2C_q^+\sim$1 to 10 aF.

\subsection{Trapped molecular ions}

\begin{figure}
\includegraphics[width=3.4in]{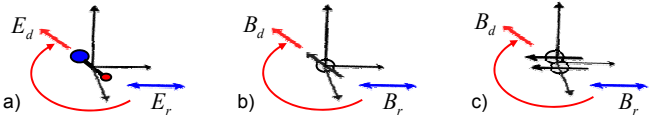}
\caption{\label{fig:vectors} Orientation of vector qubits driven by rotating fields. (a) trapped molecular ion qubits, and (b) electron spin qubits driven by a rotating field $E_d$ or $B_d$, respectively, whose rotation plane contains the resonator mode field direction; in both cases the dipole orientation follows the modulation field. In (c), the NV center spin triplet is driven by a rotating magnetic field $B_d$ whose rotation plane contains the resonator mode field ($B_r$) direction; the orientation (quantization axis) of the NV center is determined by the crystal, and is also contained in the modulation field's rotation plane.}
\end{figure}

Figure~\ref{fig:examples}(d): The trapped polar molecular ions of ref.~\onlinecite{molions} have a vector dipole moment, associated here primarily with the $J=1$ excited molecular rotational manifold with sublevels $m_J=-1,0,1$ which are degenerate at zero electric field. An electric field shifts the $J=1,m_J=\pm1$ levels down, so that a qubit can be realized with the $J=0,m_J=0$ and $J=1,m_J=0$ states. We therefore cannot simply oscillate the electric field through zero, since the $m_J=0$ state will undergo Majorana-like transitions to the $m_J=\pm1$ states near zero field (equivalently, the induced dipole's orientation will not ``follow" the applied field). Instead, we can use a \textit{rotating} electric field $E_d$ with angular frequency $\omega_m$, whose plane of rotation contains the resonator mode field axis as shown in fig.~\ref{fig:vectors}(a). In this case, as long as the rotation is not too fast, the molecular dipole will follow it, resulting in an oscillating projection of the dipole along the resonator mode field. The effect of the rotation can be expressed via Larmor's theorem as an effective magnetic field along the rotation axis: $B_{rot}=\hbar\omega_m/\gamma_{J=1}$ where $\gamma_{J=1}$ is the gyromagnetic ratio of the $J=1$ manifold. As long as the rotation is turned on and off slowly, and we restrict ourselves to $\hbar\omega_m\lesssim E_{m_J=0}-E_{m_J=\pm1}$, the states will transform adiabatically.

With this in mind, we take $E_d=2.5$ kV/cm, corresponding to a differential stark shift between the $J=0,m_J=0$ and $J=1,m_J=0$ rotational states from their zero-field splitting of 9 GHz to $\sim$13 GHz, and a splitting between $J=1,m_J=0$ and $J=1,m_J=\pm1$ of $\sim$3 GHz; we take a distance $l=1\mu$m between the qubits and modulation electrodes, so that $V_b\sim E_dl$.

Finally, although the rotation at constant field magnitude $E_d$ does not couple the $J=0,m_J=0$ to the $J=1$ manifold, there is a residual direct Jaynes-Cummings type coupling between the resonator and the $J=0\leftrightarrow J=1$ transition, given by: $\hbar g\approx(d_{01}/l)\sqrt{\hbar Y_r}/C_r=0.4$ MHz ($d_{01}\approx2.5\times10^{-29}$C$\cdot$m is the transition dipole between $J=0$ and $J=1$); this produces a negligible effect at the large qubit-resonator detunings $\gtrsim$10 GHz considered here.

\subsection{Donor electron spins in Si}

Figure~\ref{fig:examples}(e): Similar to the case of quantum dots and molecular ions which are biased with a voltage or electric field rather than a charge, an electron spin is biased with a magnetic field rather than a flux. In this case, we have: $\eta_0^-\equiv (dE^-/dI_q)/(2\sqrt{\hbar Z_r})$, $L_q^-\equiv d^2E^-/dI_q^2$, and $\delta\omega_r^-/\omega_r\approx-L_q^-/(2L_r)$. We take a circular loop of diameter $d=50$nm connected to the resonator, with the spin at its center, so that the field at the spin is given by: $B_d=\mu_0I_q/d$. As in the case of molecular ions, a rotating field (turned on and off slowly) must be used to avoid Majorana transitions between spin orientations [fig.~\ref{fig:vectors}(b)], whose rotation frequency in this case should be less than the Larmor frequency $\omega_L=\gamma_eB_d/\hbar$. For our modulation at 1 GHz, we then select a field amplitude of $B_d=1000$G, corresponding to $\omega_L/2\pi=2.8$ GHz. At this field, the hyperfine splitting of $\approx 117$ MHz \cite{Sidonor} produces a negligible $L_q^-\approx 10^{-22}$H. Our chosen parameters $Q=10^7$, $d=50$ nm, and $B_d=1000$G, while arguably not completely implausible, are admittedly extreme. The very weak interaction with a single spin dictates that such parameters are required for favorable gate parameters. One way to relax these requirements to some extent would be to implement a multiturn coil (it would need to be at the $\sim100$nm scale) to increase the coupling, or to use an ensemble of spins as a qubit.

\subsection{Nitrogen vacancy centers in diamond}

Figure~\ref{fig:examples}(f): This case is similar to the electron spin just discussed, except that the system is a spin triplet with $S=1$, and the $m_S=0$ state is shifted relative to the $m_S=\pm1$ states due to crystal-field and magnetic-dipole interactions by 2.88 GHz \cite{wrachtrup}. The axis of this internal field is fixed by the crystal, so that the alignment of the states cannot follow the external modulation field direction for weak fields. However, if we align the resonator mode field along the NV center's crystal axis, and use a modulation field rotation plane that also contains this axis [fig,~\ref{fig:vectors}(c)], we can still realize the desired effect (we take as above $B_d=1000$G). When the modulation field $B_d$ is along the crystal axis, the resonator field produces a linear Zeeman shift of the $m_S=\pm1$ states; when the modulation field is perpendicular to the axis, it mixes all three sublevels, producing a large enough avoided crossing that the $m_S=\pm1$ states can be nearly adiabatically transformed into each other by the modulation, and we can use them as our two qubit states (Note that in experiments, the $S=0,m_s=0$ state is typically used as $|g\rangle$). As above, this can be fully accounted for using an effective field associated with the rotation. As long as the modulation frequency is not too large, and the modulation is as above turned on and off slowly, nonadiabatic transitions can be neglected at the error levels of interest here.

\section{Comparison with trapped-ion gates}\label{a:ioncomp}

In laser cooling and manipulation of trapped atomic ions, the internal states of the ions are coupled to their center-of-mass motion by the photon recoil momentum $\hbar k$ associated with the (state-dependent) absorption of a photon with wavevector $k$. This coupling occurs in one of two ways: (i) via Rayleigh scattering of laser photons tuned near resonance with an electronic transition from an incident laser wavevector ${\bf k}_L$ to ${\bf k}^\prime$, which imparts a recoil momentum $\hbar ({\bf k}_L-{\bf k}^\prime)$ (radiation pressure) and can be used to implement dissipative laser cooling of the atomic motion \cite{cooling}; or (ii) via coherent (stimulated) scattering far from resonance, also known as the AC Stark shift or light shift, which can be used to realize essentially non-dissipative forces with internal-state-dependence \cite{dipole,Znature,haljan}. Our proposal is the analog of the latter case \cite{casei}.

The method we have presented, although broadly similar to the so-called M{\o}lmer-S{\o}rensen \cite{MSPRL,MSPRA} and ``ZZ" gates \cite{Znature} used for trapped ions, has some important and favorable differences from those gates. For example, in our proposal, the resonator modes are nearly (though not completely) independent of the qubits, so that adding more qubits does not generate additional nearby collective modes which must be avoided as in the case of trapped ions. Also, in the ion case the equivalent controlled-phase interaction requires the ions to be in magnetic field sensitive states, which produces inevitable dephasing \cite{Znature}; in our proposal, although the states used must also be field-sensitive, the sign of their sensitivity is oscillating at a high frequency so that to leading order dephasing is nearly absent (except in qubits with no degeneracy point). Although alternative methods for ions have been demonstrated that use states without field sensitivity \cite{MSPRL,*MSPRA,haljan}, they have additional complications which limit the gate speed, fidelity, and number of ions that can be entangled \cite{MSPRL,*MSPRA}. Finally, although the schemes used for ions are nominally insensitive to the state of the resonator, they still require a small Lamb-Dicke parameter, meaning that the ions must be localized to a much smaller region than the laser wavelength to avoid sampling a spatially dependent force \cite{MSPRL,*MSPRA}. This can be particularly challenging in the context of the ubiquitous heating observed in ion traps, which causes $\eta$ to increase in time, in the absence of active laser cooling \cite{hite}. In our case, although $\eta$ is also a small parameter (in the expansion of $E_q(p)$), this expansion does not break down as in the atomic case; in fact, most of the error sources we have discussed actually \textit{decrease} with larger $\eta$. In the expansion of eq.~\ref{eq:expand}, we included terms up to second order in the resonator displacement $\delta Q_r$, and for the parameters in table ~\ref{tab:sample} the errors associated with the second order term [c.f., eqs.~\ref{eq:epsdalpha},\ref{eq:nthdephase}] are already small enough for low error rates, and the higher-order terms can almost always be neglected completely. One possible exception to this would be associated with the presence of a strong \textit{quartic} term in the qubit energy; this results in a modulation of $\omega_r$ at \textit{twice} the input modulation frequency, which then \textit{parametrically} excites the oscillator. We have simulated this effect, and it is negligble for the parameters considered here, though it is possible it could become important in some cases.

\bibliography{longit}

\begin{thebibliography}{94}
\bibitem [{div()}]{dividers}%
  \BibitemOpen
  \href@noop {} {}\bibinfo {note} {Note that in (b) we have used inductive
  dividers (rather than mutual inductances) to retain exact duality with the
  capacitive circuit of (a); however, mutual inductances can also be used with
  straightforward modifications to the results presented here.}\BibitemShut
  {Stop}%
\bibitem [{crp()}]{crprime}%
  \BibitemOpen
  \href@noop {} {}\bibinfo {note} {Note that $C_b$ does not contribute to
  $C_r^\prime$ because the qubit voltage $V_q$ is effectively
  fixed.}\BibitemShut {Stop}%
\bibitem [{qub()}]{qubitcalc}%
  \BibitemOpen
  \href@noop {} {}\bibinfo {note} {For the flux qubit calculations, we use the
  numerical methods described in ref.~\onlinecite{MRFS}.}\BibitemShut {Stop}%
\end{thebibliography}

\end{document}